\documentclass{llncs}
\usepackage{graphicx,times,amsmath} 

\begin{document}

\title{Minimization of Quadratic Binary Functional with 
Additive Connection Matrix}
\author{Leonid Litinskii}
\institute{Center of  Optical Neural Technologies 
Scientific Research Institute for System Analysis,\\ Russian 
Academy 
of Sciences,\\ 
   44/2 Vavilov Str, Moscow 119333, Russia.\\
\email{litin@mail.ru}
}

\maketitle

\begin{abstract}
$(N \times N)$-matrix is called additive when its elements are 
pair-wise sums of $N$ real numbers $a_i$. For a quadratic binary functional  
with an additive connection matrix we succeeded in finding  
the global minimum expressing it through external parameters of the problem. 
Computer simulations show that energy surface of a quadratic binary functional 
with an additive matrix is complicate enough.
\end{abstract}

\section{Introduction}
In the present paper we analyze the classic problem of discrete mathematics 
that is minimization of a quadratic functional depending on the great number $N$
of binary variables $s_i $:
$$ E({\bf s})=-\frac {({\bf Js,s})}{2N}=-\frac{1}{2N}\sum_{i,j=1}^N 
J_{ij}s_is_j\longrightarrow\min, \ s_i=\pm 1. \eqno(1)$$
This problem arises in a lot of scientific fields of knowledge beginning 
from physics of magnetic materials and neural networks up to analysis of 
results of physical experiments and logistics. Usually the connection matrix 
${\bf{J}} = (J_{ij} )_1^N $  is supposed to be symmetric one with zero 
diagonal elements: $J_{ij}=J_{ji}, J_{ii}=0$. 
The state of the system as a whole is given by $N$-dimensional vector 
${\bf{s}} = (s_1 ,s_2 ,...,s_N )$. Such vectors will be called {\it configuration 
vectors} or simply {\it configurations}. The characteristic
$E({\bf{s}})$ that has to be minimized will be called {\it the energy} of the 
state, and the configuration providing the global minimum of the functional (1) 
will be called {\it the ground state}.
 
	In general the number of local minima of the functional (1) is exponentially large. Practically all minimization 
algorithms guarantee finding of a local minimum only. The exceptions are very rare and, as a rule, they are relied on 
specific properties of the connection matrix \cite{1}, \cite{2}. The most widespread is the {\it random 
minimization} \cite{1}. According this algorithm the spin dynamics is started from a random configuration. In randomized 
order the states of dissatisfied spins are changed. As a result the dynamic system step by step falls into the nearest local 
minimum. We used just the random minimization in our computer simulations (Section 3). 

    	Very little is known about properties of the energy surface of the functional (1), namely, about the number and the 
structure of the set of local minima, about the ground state and the probability to find it and so on. In fact there is only one 
nontrivial connection matrix for which the ground state of the functional (1) can be indicated exactly. This is the Hebb 
matrix in the case when the value of the loading parameter $\alpha  = M/N$
     is small: $\alpha  < 0.07$
     \cite{3}. Then the global minimum of the functional (1) is achieved at any of $M$
     random patterns. 

	Due to discrete character of the problem its theoretical analysis is very rare. From recent results let us point out the 
papers \cite{4}, \cite{5}, where the authors succeeded in connecting the depth of the local minimum with the probability of 
its random finding, and also described some characteristics of the energy surface. 

    In our work we introduce a class of {\it additive matrices} whose elements are pair-wise sums of a set of 
predetermined numbers $a_i $:
$$J_{ij}=(1-\delta_{ij})(a_i+a_j),\quad i,j= 1,...,N,\quad{\rm{where}}\quad
\left\{{a_i}\right\}_1^N\in{\bf{R}}^{\bf{1}} 
\eqno(2)$$
and $\delta_{ij}$ is the Kronecker delta symbol.

The additive matrices generalize a special class of Hebb's matrices analyzed in \cite{6}. For the functional (1) with the 
connection matrix (2) the ground state can be obtained exactly. We succeeded in presentation additive matrices in the form 
when the dependence of the ground state on external parameters of the problem can be described analytically. When the 
ground state is known, interesting results can be obtained with the aid of computer simulation. In the next Section we 
present the theory relating to the problem. In Section 3 we give the results of computer simulations. 

\section{The Ground State}

{\bf 1.} It can be verified directly that for an additive matrix (2) the value of the functional (1) is equal to 
$$E({\bf{s}}) =  \frac{({\bf e},{\bf a})-({\bf s},{\bf e})({\bf s},{\bf a})}{N}.\eqno(3)$$
Here ${\bf{a}} = (a_1 ,a_2 ,..,a_N )$
is $N$-dimensional vector whose coordinates are $a_i $, 
and ${\bf{e}} = (1,1,...,1)$ is the "bisector" of the principal 
orthant of the space ${\bf{R}}^{\bf{N}}$. 
From minimization of the functional (3) one can pass to {\it maximization} of the functional 
		
$$ F({\bf s})=({\bf s},{\bf e})({\bf s},{\bf a})\longrightarrow\max.\eqno(4)$$												

Let us denote by $\Sigma _k $ {\it the class} of all configurations $\bf s$ for which exactly $k$ 
coordinates are equal "-1":
$$\Sigma _k  = \left\{ {{\bf{s}}:\;\left( {{\bf{s}},{\bf{e}}} \right) = N - 2k} \right\},\;\;k = 0,1,...,N.$$
The class $\Sigma _k $ consists of $C_k^N$
configurations. For all these configurations the first multiplier in the expression (4) takes the
same value $N - 2k$. Consequently, to maximize (4) {\it among configurations from the class} $\Sigma _k $
, it is sufficient to find a vector ${\bf{s}} \in \Sigma _k$ maximizing the 
scalar product $({\bf{s}},{\bf{a}})$. 
This problem is not so difficult (see item 3). 

{\bf 2.} Suppose, we can find the vector ${\bf{s}}$
maximizing the scalar product $({\bf{s}},{\bf{a}})$
in the class $\Sigma _k$. Let us denote this vector as ${\bf{s}}(k)$, 
and let the value of the functional (4) for this vector be 
$F_k  = F({\bf{s}}(k))$: 
$$F_k  = (N - 2k)({\bf{s}}(k),{\bf{a}}) = \mathop {\max }\limits_{{\bf{s}} \in \Sigma _k } F({\bf{s}}),\;\;k = 
0,1,..,N.$$
When finding all these vectors ${\bf{s}}(k)$ ($k = 0,1,..,N$), it is easy to find the global maximum of the functional (4), since the functional reaches its maximal value on one of the vectors ${\bf{s}}(k)$.	

Note we do not need to compare between themselves all $N+1$ numbers $F_k$, 
but the first half of them only. 
The reason is that for any $k$ the classes $\Sigma_k$
and $\Sigma_{N-k}$ are inversion of each other: 
$\Sigma_{N-k}=-\Sigma_k$. Since for any configuration ${\bf{s}}$
 the equality $F({\bf{s}})=F(-{\bf{s}})$
 is fulfilled, we obtain that $F_k  = F_{N - k} $ for all values of $k$. 
Combining the cases of even and odd $N$
     in one formula we obtain that to find the global maximum of the functional 
(4) it is necessary to find the largest of the 
values $F_k $, when $k \le n = \left[ {{N \mathord{\left/{\vphantom {N 2}} 
\right.\kern-\nulldelimiterspace} 2}} \right]$: 
$$F_0 ,\;F_1 ,..,\;F_n ,\;\;n = \left[ {\frac{N}{2}} \right].$$
		
{\bf 3.} Without loss of generality the numbers $\left\{ {a_i } \right\}_1^N $
 can be put in order according their increase: 
    $$a_1  < a_2  < ... < a_N .\eqno(5)$$
Let us take any $k \le n$. It is easy to see that the scalar product 
$\left( {{\bf{s}},{\bf{a}}} \right)$ reaches its maximum inside the 
class $\Sigma _k$ when the configuration vector is 
$${\bf{s}}(k) = (\underbrace { - 1, - 1,..., - 1,}_k\;1,\;1,..\;1).\eqno(6)$$
Indeed,
$$\left({{\bf{s}}(k),{\bf{a}}} \right) = -\sum\limits_{i=1}^k a_i + \sum\limits_{i=k+1}^N a_i= 
\hat a - 2\hat a_k,$$ 
where 
$$\hat a=\sum\limits_{i = 1}^N {a_i },\mbox{  } \hat a_k=\sum\limits_{i = 1}^k a_i, 
\mbox{  and  } \hat a_0=0.\eqno(7)$$

Let ${\bf{s}}$ be another configuration vector from the class $\Sigma _k $
for which numbers of negative coordinates $j_1  < j_2  < ... < j_k $
 dose not take the first places. The scalar product $\left( {{\bf{s}},{\bf{a}}} \right)$
 is equal to 
$\left({{\bf{s}},{\bf{a}}}\right)=\hat a-2\sum\limits_{i=1}^k{a_{j_i }}$, 
and inequality 
$\left( {{\bf{s}}(k),{\bf{a}}} \right) > \left( {{\bf{s}},{\bf{a}}} \right)$
 is fulfilled since 
$\sum\limits_{i = 1}^k {a_i }  < \sum\limits_{i = 1}^k {a_{j_i } } $
for any set of indices $\{ j_1 ,j_2 ,...,j_k \}$
 that differs from $\{ 1,2,...,k\} $.

    Thus, under the condition of ordering (5), to find the global minimum of 
the functional (3) it is necessary to find the largest among the numbers 
$$F_k  = (N - 2k)\left({\hat a - 2\hat a_k}\right), k =0,1,..,n=\left[\frac{N}2\right],\eqno(8)$$
where $\hat a$ and $\hat a_k$ are given in Eq.(7).

{\bf 4.} The initial problem (1)-(2) can be considered as solved: the expressions (6) restrict the set of configurations among which 
the ground state of the functional (2) should be found. To define which configuration is the ground state it is necessary to 
calculate $n$
 numbers (8) and find the largest among them. It reminds unclear under which conditions this or that configuration (6) 
would be the ground state. If any of them will be the ground state or not? It turned out that these questions can be answered. 

    Without loss of generality let us suppose that the numbers $a_i $
     have a special form:
$$a_i=\alpha_i-t,\quad \alpha _i  \in [0,1],\ t \ge 0.\eqno(9)$$
In this presentation the values $\alpha _i $
 are positive numbers from the unit interval, and the positive parameter $t$
can take an arbitrary value. It is not difficult to see that from the point of view of minimization of our 
functional an arbitrary set of numbers $a_i$ can be reduced to the form (9).
For example, let us suppose that $a_1  < ... < a_N  < 0$ and $a_N  - a_1  
\le 1$. Then we set $\alpha _i  = a_i  - a_1 $ and $t = \left| {a_1 } \right|$. 
This means that the numbers $a_i $ have the form (9). 
On the contrary, let the initial numbers $\tilde a_i$ have different signs 
and take on arbitrary values:
$\tilde a_1  < ...0 < ... < \tilde a_N $, and
$\tilde a = \max \left( {\left| {\tilde a_1 } \right|,\tilde a_N } \right) 
>>1$. Let us normalize these numbers dividing them by $2\tilde a$: 
$a_i  = \tilde a_i/{2\tilde a} \in 
\left[ -1/2,+1/2\right]$. 
It is clear that the solution of the problem (1) is the same when 
we use initial numbers $\tilde a_i $ or normalized numbers $a_i $. 
The last numbers can be presented in the form (9), if we set 
$\alpha_i  = a_i  + \frac{1}{2}$ and $t = \frac{1}{2}$. 
From our argumentation it follows that the numbers $a_i $ can always 
be presented in the form (9). Then the following statement is right 
(the proof see in the Appendix). 

{\bf Theorem.} When $t$
increasing from the initial value $t = 0$, the ground state sequentially 
coincides with the vectors ${\bf{s}}(k)$
(6) in the following order: 
$${\bf{s}}(0)\to{\bf{s}}(1)\to ...\to{\bf{s}}(k - 1)\to{\bf{s}}(k)\to ...
\to{\bf{s}}\left({n - 1}\right)\to{\bf{s}}\left(n\right).\eqno(10)$$
The jump of the ground state ${\bf s}(k - 1)\to{\bf s}(k)$
occurs when $t$ transfers through the critical value: 
$$t_k  = \frac{\hat\alpha-2\hat\alpha_k+(N-2k+2)\alpha_k}{2(N-2k+1)},
\quad k =1,2..,n,\eqno(11)$$
where analogously of Eq.(7) $\hat\alpha=\sum\limits_{i=1}^N\alpha_i$ and 
$\hat\alpha_k=\sum\limits_{i=1}^k\alpha_i$.
When $t$ belongs to the interval $[t_k ,t_{k + 1} ]$, the ground state 
of the functional is the configuration ${\bf{s}}(k)$. 

This theorem generalizes the previous results obtained in \cite{6}. The theorem 
describes exhaustively the behavior of 
the ground state for the problem (1)-(2). Depending on the values of  
external parameters $\{a_i\}$ each of the 
configurations ${\bf{s}}(k)$, $k = 0,1,...,n$
 can turn out to be the ground state of the functional. For $t$
     from the interval $[t_k ,t_{k + 1} ]$
     the energy of 
the ground state is the linear function of the parameter $t$. 
It can be easily seen from the expressions (8) substituting the values $a_i$
in the form (9): 
$$E_k (t) = A_k  + t\cdot B_k ,\;\;k = 0,1,...,n,\eqno(12)$$
where up to the factor $1/{2N}$ we have:
$$A_k=\hat\alpha-(N-2k)(\hat\alpha-2\hat\alpha_k),\;\;B_k=(N-2k)^2-N.\eqno(13)$$

\begin{figure}[htp]
\centering
\includegraphics[width=4in]{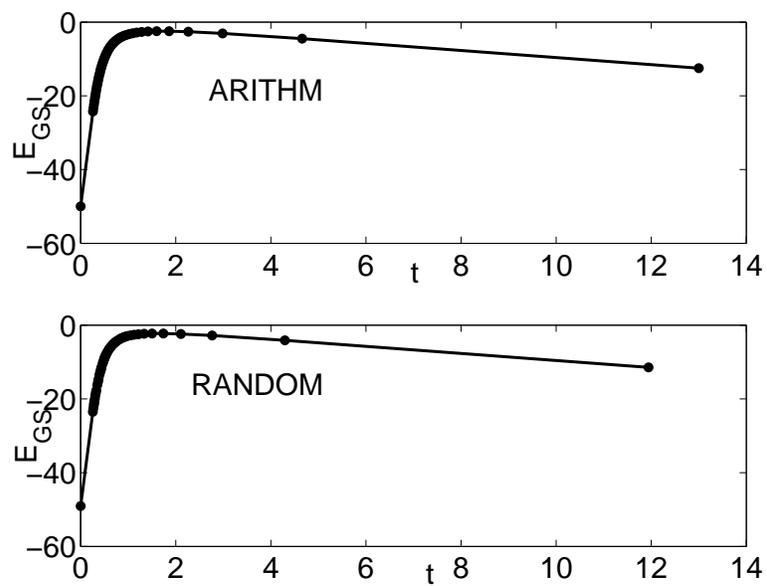}
\caption{The dependence of the ground state energy $E_{GS}$ on the parameter 
$t$ for additive matrices of the dimensionality $N=100$: the upper one is 
the arithmetical additive matrix, and the lower one is 
the random matrix (see the body of the text).
The values $E_{GS}$ for the points $t_k$ are marked.}
\end{figure}
                                                                     			
In Fig.1 for $N = 100$ it is shown how the energy of the ground state 
depends on the parameter $t$. The upper panel corresponds to the case when 
the values $\alpha _i $ constitute the arithmetical progression: 
$\alpha _i  = i/N,\;\;i = 1,2,..,N$. On the lower panel the analogous 
plot is shown for a random additive matrix, when $\alpha _i $
     are random numbers from the interval [0, 1]. Along the abscissa axis 
the values of the parameter $t$ are shown, along the axis of ordinates we 
show the energy of the ground state calculated in the points $t_k $ (11). 
The first value of the parameter $t$ for which the energy of the ground 
state is calculated is equal to zero: $t_0  = 0$. Since both plots are very 
similar, we analyze only one of them; for example the upper one. 

    	We see that the energy of the ground state is nontrivially depended 
on the parameter $t$. For small values, $t\sim 0$, very deep minima correspond 
to the ground state. Then, when $t$ increases, the depth of the global minimum 
decreases very quickly and it reaches a minimal value when $t \approx 2$. 
For these values of $t$ all matrix elements become negative. 
During further increase of $t$ the depth of the global minimum 
slowly but steadily increases. It becomes deeper and deeper.
Which properties of the energy surface reflect non-monotone change of the 
depth of the global minimum?  What properties are responsible for its 
minimal depth? For the time being we cannot answer these questions. 
Using formulae (10)-(13) everyone can be certain of 
universal character of the curves shown in Fig.1.

    Up till now we can neither extend these results onto local minima of the functional, nor obtain analytical description 
of other interesting characteristics such as the number of different minima, 
distribution of local minima with respect to their 
depths and distances to the ground state and so on. However, if the ground 
state is known, these characteristics can be 
studied with the aid of computer simulations. Now we turn to presentation 
of these results. 

\section{Computer Simulation} 

    For given $
    N
    $
     and $\left\{ {\alpha _i } \right\}$
     for each value of $
    t
    $
     we generated an additive matrix. We did $10^5 $
    random starts (see Introduction) and obtained the same number of local minima. For each minimum we fixed its depth 
(the energy$E_l $
    ), the relative Hamming distance $D_l $
     between the minimum and the ground state and other characteristics. Thus as a result of a great number of random 
trials for each value of $
    t
    $
     we could estimate: a) the probability of random finding of the ground state $p_{GS} $
    ; b) the deepest of the obtained minimum and the distance from it to the ground state; c) the number of different minima 
$K$
    , their distribution over energies and distances from the ground state and so on.  The parameter $
    t
    $
     was varied from zero up to the maximal value $t_n $
    . For two dimensionalities $N = 100$ and $N = 1000$ such experiments were 
done for both arithmetical and random additive matrices.  

\begin{figure}[htp]
\centering
\includegraphics[width=4in]{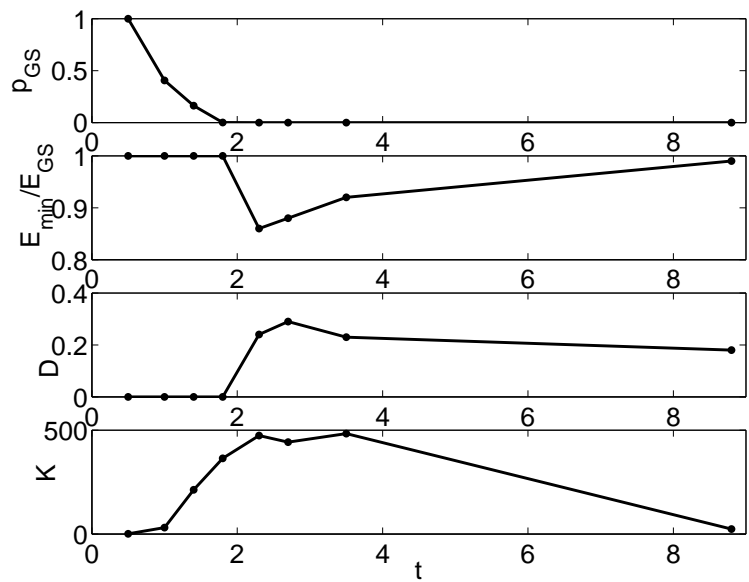}
\caption{For arithmetical additive matrix of dimensionality $N=100$ 
the following graphs are shown: on the upper panel is the probability 
to find the ground state; on the next panel is the ratio of depth of 
the deepest found minimum to the depth of the global minimum; 
on the next panel is the relative Hamming distance between the 
deepest minimum and the ground state; 
on the bottom panel is the number of different energies of local minima.}
\end{figure}
            
In Fig.2 for the arithmetical additive 
matrix of dimensionality $N = 100$ the dependence of some of the 
listed characteristics on the parameter $t$ is shown. 
Let us explain what the graphs shown on different panels of the figure mean.
    
{\bf On the upper panel} the probability to find the ground state $p_{GS}$
is shown. We see that in the region of small values of $t$
($t < 1.8$), where the depth of the global minimum is large, 
the probability to find the ground state is notably different from zero. 
On the contrary, in the region of large values of $t$, where the global 
minimum becomes rather shallow, the probability to find 
it is equal to zero (it is less 
than $10^{-5}$). 
At the same time it is not important which  
configuration ${\bf{s}}(k)$ is the ground state.

    	Apparently, such behavior of the probability $p_{GS} $
     is one more confirmation of the law, which was theoretically predicted in \cite{3}, \cite{4}: the deeper minimum, the 
greater probability to find it under the random search. 

For the matrix of dimensionality $N = 1000$
the behavior of the given characteristic is an analogous one. The value of the parameter 
$t$ for which the probability to find the ground state becomes zero, increases up to the value 
$t \approx 4$. 

	{\bf On the second panel from the top} the ratio of the found deepest minimum $E_{\min } $
 to the global minimum $E_{GS} $
, $E_{\min } /E_{GS} $
, is shown.  This ratio takes on a value from the interval [0, 1].  At first, while the ground state still can be found, this ratio 
is equal to 1. Then in the region of the values $t \approx 2.3 - 2.5$
 this characteristic has a sharp downward excursion, which soon changes to a steady increasing and tends to 1 
asymptotically. The minimal value of this characteristic is 
$E_{\min } /E_{GS}  \approx 0.85$. 
It shows that in the worst case the objective function is 15\%  
less than the optimal value.

	For matrices of dimensionality $N = 1000$
 the behavior of the ratio $E_{\min } /E_{GS} $
 is absolutely analogous. The deepest downward excursion of the graph takes place when $t \approx 4$
, and its depth increases noticeably: the minimal value of the ratio is equal 
$E_{\min } /E_{GS}  \approx 0.5$. In other words, when the dimensionality 
of the problem increases the found suboptimal solution will be worse comparing 
with the global minimum.  

	Note, that for the large values of $t\sim 7-8$
, when the ratio $E_{\min } /E_{GS} $
 is close to 1, the probability to find the ground state as before is equal to 0. The same also takes place in the case $N = 
1000$.  

	{\bf On the second panel from the bottom} it is shown how the distance $
D
$
between the deepest local minimum and the ground state depends on the 
parameter $t$. (By the distance we understand the relative Hamming distance 
$D = {{\left( {N - abs({\bf{s}},{\bf{s}}')} \right)}\mathord{\left/
{\vphantom {{\left( {N - abs({\bf{s}},{\bf{s}}')} \right)} {2N}}} \right.
 \kern-\nulldelimiterspace} {2N}} \in [0,0.5]$.) 

At first, while the ground state still can be found this distance is equal to 0 (see the beginning of the graph). Then in the 
interval of the "worst" values of $
t
$
 the distance $
D
$
 increases sharply up to the value $D_{\max }  \approx 0.3$
. After that the distance between the deepest local minimum and the ground state is stabilized near the value $
D = 0.2
$. Let us add that for additive matrices of dimensionality $
N = 1000
$
 suboptimal solution is far away from the ground state. This distance is $
D \approx 0.4
$
. 

    The general conclusion is as follows: for rather large values of $
t
$
, when as a result of the random search it is possible to find suboptimal solution only, this solution is sufficiently far from 
the ground state. However, the ratio of the minima depths $E_{\min } /E_{GS} $
 can be of order of 1 (in Fig.2 this situation corresponds to the values of $
t > 4
$
). This combination of properties  is possible only if the energy surface consists of a large number of local minima, which 
depths not strongly differ one from each other and from the global minimum. We may conclude, that for large values of $
t
$
, when elements of connection matrix are large negative numbers, the construction  of the energy surface is as aforesaid. 

{\bf On the bottom panel} we show the dependence of the number of 
different energies of local minima $K$ on the value of the parameter $t$. 
As a rule each energy is many times degenerated. To estimate the number of different local minima it is 
necessary to analyze how many different configurations correspond to the same energy.  Nevertheless, such characteristic 
as the number of energy levels is also of interest. 

For an arithmetical additive matrix of the dimensionality $N = 100$
 the maximal value of the characteristic $K$
is reached in the region $t\sim 3$. This maximum is comparatively small,
$\sim 500$.  However, it turns out that each energy is many times 
degenerated, and the number of different local minima is an orders of 
magnitude greater. For a random additive matrix of 
the same dimensionality the maximal value of the characteristic 
$ K$ is equal to tens of thousands (the graph is not presented). 

For the additive matrices of the dimensionality 
$N = 1000$ the general form of the graph of the characteristic 
$K$ is analogous. In this case the maximal value, $K_{\max }  \sim 4\cdot 10^4 $, 
is reached in the region $t \approx 5$. Since for each 
$t$ only $10^5 $ random starts have been done, this means that each second 
start leads the system into new local minimum. In other 
words, for middle values of  $t$ the number of local minima is very big. 

\section{Discussion and Conclusions} 

For additive matrices the method of finding of the global minimum of 
the quadratic binary functional is 
pointed out. We propose the $t$
-parametrization of additive matrices that allows one to get an exhaustive classification for all variants possible for the 
ground state. 

     For not great values of $t$
 (let us say for $t \in [0,2]$
) among matrix elements there are positive as well as negative ones; or all elements are negative, but they are small in 
modulus. In this case the depth of the global minimum is very big.  
Here the probability to find the ground state in random search is rather high: 
$p_{GS}\sim 0.5-1.0$.
It can be supposed that in this case the energy surface has a small number of local minima whose depths noticeably less 
then the depth of the global minimum. 

On the contrary, for the great values of $t$
 all matrix elements are negative and they are big  in modulus. In this case it is practically impossible to find the ground 
state with the aid of the random minimization, since the probability to get into the global minimum is negligible small. 
Apparently in this case the energy surface contains very large number of local minima that only slightly differ from each 
other in depths. Here the global minimum is only insignificantly deeper 
than local minima. Varying the value of the parameter $t$
 it is possible to get over from one type of the energy surface to the other 
one. So, additive matrices are good models for 
examining the energy surfaces in the general case.  

	By this time additive matrices for large values of $t$
can be used for testing of new algorithms of the quadratic binary minimization. 
Indeed, on the one hand, with the aid of the formulae (10)-(13) the ground 
state always can be found. On the other hand, for large values of the 
parameter $t$ it is practically impossible to find the ground state with 
the aid of the random minimization.

\section*{Acknowledgment}
The work has been done with participation of Anton Pichugin 
in the framework of the program supported by the grant \#356.2008.9 
of President of Russian Federation and in part 
by Russian Basic Research Foundation (grant \#09-07-00159).

\section*{Appendix}
In the beginning of Section 2 it was shown that only one of configuration vectors ${\bf{s}}(k)$
(6), $k = 0,1,...,n = \left[N/2\right]$ can be the ground state. Using the representation (9) of $a_i $
it is easy to obtain Eq. (12) for the energies of  ${\bf{s}}(k)$-configurations: $E_k (t) = A_k  + tB_k $, where $A_k $ and $B_k $ are given by Eq.(13). As functions of the parameter $t$ energies $E_k (t)$ are straight lines. We have to analyze the behavior of the set $\left\{ {E_k (t)} \right\}_0^n $. When a straight line $E_l (t)$
{\it  is lower all other straight lines, the configuration ${\bf{s}}(l)$
is the ground state}. 

For simplicity we restrict ourselves to the case of even$N = 2n$. Let us write down the expression (13) in more details:
$$
\begin{array}{*{20}c}
   {\begin{array}{*{20}c}
   {A_0  =  - (N - 1)\hat \alpha } &  <  & {A_1 } & {\begin{array}{*{20}c}
    <  & {...} &  <  & {A_n  = \hat \alpha ,}  \\
\end{array}}  \\
\end{array}}  \\
   {\begin{array}{*{20}c}
   {B_0  = (N - 1)N} &  >  & {B_1 } & {\begin{array}{*{20}c}
    >  & {...} &  >  & {B_n  =  - N.}  \\
\end{array}}  \\
\end{array}}  \\
\end{array}\eqno(A1)
$$

When $k$ increasing, the free term $A_k $ of the straight line $E_k (t)$ 
increases monotonically. In other words, the intersection of the straight 
line with ordinate axis rises higher and higher. On the other hand, when $k$
increasing the coefficient $B_k $ decreases monotonically, so that in the 
end it even becomes negative.  For the case $N = 6$ the typical behavior 
of the set of straight lines $\left\{ {E_k (t)} \right\}_0^n $ is shown 
in Fig.3. We use this figure to explain how the ground state depends on 
the parameter $t$.  

When $t = 0$ all the matrix elements are positive and the configuration 
${\bf{s}}(0) = (1,...1)$ is the ground state. Let us increase $t$ little by 
little. At first the straight line $E_0 (t)$  is lower than all other 
straight lines. Consequently, ${\bf{s}}(0)$  remains the ground state.  
Than for some value of the parameter $t$ the straight line $E_0 (t)$ 
is intersected by another straight line. After that this straight line 
turns out to be lower than all other straight lines. Taking into account 
the relations (A1) it is easy to see that the first straight line that 
intersects 
$E_0 (t)$ is $E_1 (t)$ (see also Fig.3). After this intersection the 
configuration ${\bf{s}}(1)$ becomes the ground state.  
It is the ground state until another straight line intersects the straight 
line $E_1 (t)$. After that this straight line turns out to be lower than 
all other straight lines. From the aforesaid argumentation it is evident 
that it will be the straight line $E_2 (t)$ (see Fig.3). Then the configuration 
${\bf{s}}(2)$  will be the ground state, and so on. It can be shown that if 
the straight line $E_{k - 1} (t)$  is lower than all other straight lines, 
the first straight line that intersects $E_{k - 1} (t)$ is $E_k (t)$. 
The intersection takes place in the point $t_k $ (11) that is the solution 
of equation $A_{k - 1}+t\cdot B_{k - 1}=A_k+t\cdot B_k $.

\begin{figure}[htp]
\centering
\includegraphics[width=4in]{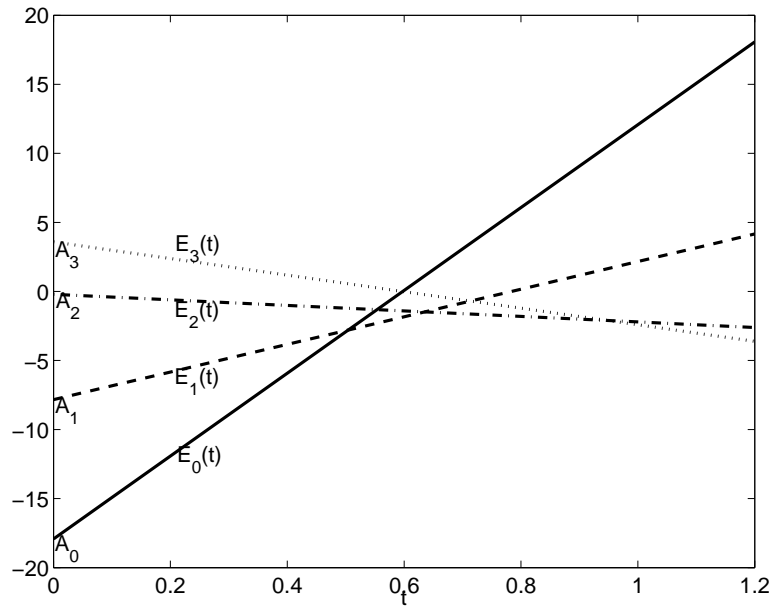}
\caption{For random additive matrix of dimensionality $N=6$ the straight lines 
$E_k(t)=A_k+t\cdot B_k$ are shown for $k=0,1,2,3$ (see the body of the text).}
\end{figure}
\end{document}